\documentstyle[preprint,eqsecnum,aps,epsfig]{revtex}

\begin{document}
\tightenlines

\preprint{
\hfill\vbox{
\hbox{\bf MADPH-00-1189}
\hbox{September 2000}}}

\title{\vspace*{.5in}
Semileptonic form factors -- a model-independent approach}

\author{Todd Coleman and M. G. Olsson}
\address{Department of Physics, University of Wisconsin, 1150
University Avenue, Madison, WI 53706}

\author{Sini\v{s}a Veseli}
\address{Fermi National Accelerator Laboratory, P.O. Box 500, Batavia,
IL 60510}

\maketitle

\vspace{.5in}

\begin{abstract}
We demonstrate that the $B\rightarrow D^{(*)}l\nu$ form factors can be
accurately predicted given the slope parameter $\rho^2$ of the 
Isgur-Wise function.
Only weak assumptions, consistent with lattice results,
on the wavefunction for the light degrees of freedom
are required to
establish this result.  We observe that the QCD and $\frac{1}{m_Q}$
corrections can be systematically represented by an effective 
Isgur-Wise function
of shifted slope.  This greatly simplifies the analysis of semileptonic
$B$ decay.
We also investigate what the available semileptonic data can tell us
about lattice QCD and Heavy Quark Effective Theory. A rigorous identity  
relating the form factor slope difference $\rho_{D^*}^2-\rho_{A_1}^2$ to a  
combination of form factor intercepts is found. The identity provides a means  
of checking theoretically evaluated intercepts with experiment.
\end{abstract}

\newpage
\section{Outline}
We obtain a nearly
model-independent description of the Isgur-Wise (IW) function \cite{Isgur},
$\xi(w)$, in terms of a single measurable parameter --- the slope at
zero-recoil, $\rho^2$.  Only modest assumptions about the shape
of the heavy-light wavefuction are required, and these
are consistent with an established
lattice result.  We obtain a simple functional
form for the IW function in terms of $\rho^2$:
\begin{equation}\label{fcnform}
\xi(w, \rho^2)=\frac{2(w+1)}{[w+1+(\rho^2-\frac{1}{2})(w-1)]^2} \,.
\end{equation}

We also demonstrate that the effect of radiative QCD corrections and
$\frac{1}{m}$ corrections can be described as the product of
a term linear
in $w$ and the
IW function.  The resulting function is also
approximately an IW function of shifted slope.  Thus, fitting the
above functional form to the measured form factors provides an accurate
description of the extrapolation to zero-recoil.
Over the range of expected slopes, our result is in
good agreement with the work of \cite{CN,CLN,BGL}
which is based on dispersion
relations.

Recent analyses \cite{CLEO,CLEOD} of $B\rightarrow Dl\nu$ and
$B\rightarrow D^*l\nu$ decays extract two physical form
factors and two ratios of physical form factors.  We show that
this information can be directly related to QCD predictions.

In Section \ref{intro} we provide some background formalism behind
form factors for semileptonic $B$-decay and the Isgur-Wise function.
We present our description of the IW function in terms of the slope
parameter, $\rho^2$, in Section~\ref{IW}.  In Section~\ref{FF}  we discuss the  
effects of corrections
to the heavy-quark limit and their equivalence to slope-shifted IW functions.  
The relationship of HQET and QCD lattice simulation to semileptonic data is  
explored in Section~\ref{QCD}.

\section{Introduction}\label{intro}
As the observed sample of semileptonic $B$ decays accumulates, the need
for a rigorous method of analysis becomes more pressing.  Even in the
heavy quark limit, a universal form factor is required for each light
degree of freedom state of the meson.  To compute the Isgur-Wise (IW)
functions $\xi(w)$, where $w=v^{\mu}v_{\mu}'$, requires considerable
knowledge of the heavy-light meson dynamics.  For this reason, the IW
function is usually thought of as ``model-dependent'' and not
susceptible to reliable calculation.  In the first portion of
this paper we seek a balance
between rigorous constraints and phenomenological usefulness.  We
will demonstrate, from heavy quark symmetry and some qualitative
lattice results, that the IW function is accurately determined by specifying
only the slope parameter, $\rho^2$.

Experimentalists commonly adopt the straightforward procedure of
expanding form factors about the meson zero-recoil point ($w=1$),
\begin{equation}\label{1-1}
\xi(w)=1-\rho^2(w-1)+c(w-1)^2+\ldots
\end{equation}
Burdman \cite{burd} has pointed out that the effect of the curvature term $c$
is significant but that, for statistical reasons, much predictive power
is lost when it
is a free parameter.  A ``model-independent'' relation between
$c$ and the slope parameter $\rho^2$ has been proposed by Boyd, Grinstein and Lebed~\cite{boyd}. This method has been modified by Caprini and
Neubert \cite{CN} and by Caprini, Lellouch, and Neubert
 in expanded form~\cite{CLN}. This result has been criticized however by Boyd,
Grinstein, and Lebed \cite{BGL}.
These latter authors propose a similar but weaker relation between
$c$ and $\rho^2$.
In Section \ref{IW}, we propose a rigorous
one-parameter expression for the IW function.

The second observation we make here concerns the relation of the IW
function to the actual (physical) form factors with $\frac{1}{m_Q}$
and QCD radiative corrections.  We show in section \ref{FF} that all
 of these
corrections can be distilled into a new effective
IW function with a shift in
slope.  The analysis of experimental data is thereby greatly simplified.
The slopes appearing in  $B\rightarrow D$ and $B\rightarrow D^*$  will be
different and this difference can be compared to the theoretical
predictions.

The decay rate for $B\rightarrow
D^{(*)}l\bar{\nu}$ can be written \cite{VO1} as,
\begin{equation}\label{1-2}
\frac{d\Gamma}{dw} = \frac{G_F^2|V_{cb}|^2}{48\pi^3}m_B^2m_{D^{(*)}}^3
 \sqrt{w^2-1}f^{(*)}|F^{(*)}|^2 \,,
\end{equation}
where for $B\rightarrow Dl\bar{\nu}$
\[ f=(w^2-1)(1-r)^2 \,, \]
\begin{equation}\label{1-3a}
r=\frac{m_D}{m_B} \approx 0.35 \,,
\end{equation}
and for $B\rightarrow D^{*}l\bar{\nu}$
\[
f^*=(w+1)[(w+1)(1-r^*)^2 + 4w(1-2wr^* + r^{*2})] \;,
\]
\begin{equation}\label{1-3}
r^*=\frac{m_{D^*}}{m_B} \approx 0.38 \;.
\end{equation}

The two form factors $F(w)$ and $F^*(w)$ can be expressed \cite{FN} in
terms of the fundamental form factors $h_+$, $h_-$, $h_V$, $h_{A_1}$,
$h_{A_2}$, and $h_{A_3}$.  These form
factors can in turn be related to the Isgur-Wise function through
\begin{equation}\label{1-4}
h_i(w)=[\alpha_i+\beta_i(w) +\gamma_i(w)]\xi(w) \;.
\end{equation}
where $\alpha_+=\alpha_V=\alpha_{A_1}=\alpha_{A_3}=1$ and $\alpha_- =
\alpha_{A_2}=0$.  The $\beta_i(w)$ describe perturbative radiative
corrections and are in principle predictable within the heavy quark
effective theory.
The $\gamma_i(w)$ represent the power ($\frac{1}{m_Q}$)
deviations from heavy quark symmetry and require further theoretical
assumptions \cite{FN}.
 In Section \ref{FF} we will point
out that,
for those form factors which do not vanish in the heavy quark limit,
the pre-factors in (\ref{1-4}) can
nevertheless be approximately absorbed into an effective
IW function which has a flavor and spin-dependent slope
\begin{equation}\label{1-5}
h_i(w) \simeq h_i(1)\xi(w,\rho_i^2)
\end{equation}

\section{Model independent parametrization of the Isgur-Wise function}
\label{IW}
The IW function appears to
require detailed information on the nature of heavy-light
dynamics.  In this section, we propose a method to distill most of this
knowledge into a single parameter --- the slope $\rho^2$.
Almost all model dependence will be seen to vanish once the slope
is specified.

In the heavy-light limit, heavy quark symmetry provides a unique
prescription \cite{biglist,VO2} relating the IW function to the wavefunction
and light degrees of freedom energy E,
\begin{equation} \label{IW1}
\xi(w)=\frac{2}{w+1}\left< j_0 \Bigl(2Er\sqrt{\frac{w-1}{w+1}}
\Bigr) \right> \,.
\end{equation}
The expectation value may involve a multi-component wavefunction such as
from the Dirac equation.  For simplicity of notation, we assume a single
component wavefunction $\psi(\mbox{$\bf{r}$})
	 = R(r)Y_{l}^{m}(\theta, \phi)$.  The
expectation value is consequently defined by
\begin{equation}
\left< A(r) \right> \equiv
    \int\limits_0^{\infty} dr \: r^2R^2(r)A(r) \,.
\end{equation}
Using $\lim_{y\ll 1}j_0(y)=1-\frac{1}{6}y^2+\ldots\,$, we may expand
(\ref{IW1}) about the zero recoil point $w=1$ to yield:
\begin{eqnarray}\label{IW3}
\xi(w) &\approx& [1-\frac{1}{2}(w-1)+\ldots]
          [<\! 1 \!> - \frac{4E^2}{6}
    (\frac{w-1}{w+1})<\!r^2\!>+\ldots] \\
\xi(w) &\approx& 1 - \rho^2(w-1)+\ldots
\end{eqnarray}
where
\begin{equation}\label{IW4}
\rho^2=\frac{1}{2} + \frac{E^2}{3} \left< r^2 \right> \;.
\end{equation}
We now use (\ref{IW4}) to eliminate the energy $E$ in the general
expression (\ref{IW1}) to obtain:
\begin{equation}\label{IW5}
\xi(w)=\frac{2}{w+1}\biggl<j_o\Bigl(\frac{r}{r_{rms}}\sqrt{\frac
	{12(\rho^2-\frac{1}{2})(w-1)}{w+1}}\Bigr)\biggr> \;,
\end{equation}
where $r_{rms}=\sqrt{\left< r^2 \right>}$.

In the above (\ref{IW5}), if we know the slope ($\rho^2$) and the radial
ground state heavy quark wavefunction, we can compute the IW function
for all $w$.  We can make the wavefunction dependence more explicit by
introducing the dimensionless quantities
\begin{eqnarray}
x &\equiv& \frac{r}{r_0} \,,\nonumber \\
R(r)&=&R_0f(\frac{r}{r_0})=R_0f(x) \,,
\end{eqnarray}
where $r_0$ is a hadronic scale factor.
We define the $n^{th}$ moment of $f^2$ as
\begin{equation}
N_n\equiv\int\limits_0^{\infty}dx \: x^{n+2} f^2(x)
\end{equation}
so that wave function normalization and $\left< r^2 \right>$ become:
\begin{eqnarray}
r_0^3R_0^2N_0 &=&1  \,, \nonumber \\
<\!r^2\!> &=& r_0^5R_0^2N_2=r_0^2\frac{N_2}{N_0} \,.
\end{eqnarray}
The IW function (\ref{IW5}) is then
\begin{equation}\label{bigIW}
\xi(w)=\frac{2}{w+1} \frac{1}{N_0} \int\limits_0^{\infty}
       dx \: x^2 f^2(x) j_0\biggl(x
   \sqrt{\frac
      {12(\rho^2-\frac{1}{2})(w-1)N_0}{(w+1)N_2}}\biggr)  \,.
\end{equation}
It should be noted that both the normalization constant $R_0$ and the
hadronic scale parameter $r_0$ do not appear in the above expression for
the IW function.
The IW function thus depends only on the dimensionless slope parameter
$\rho^2$ and one dimensionless function $f(x)$ which is essentially the
heavy-light wavefunction.

\begin{figure}[p]
\centering\leavevmode
\epsfig{file=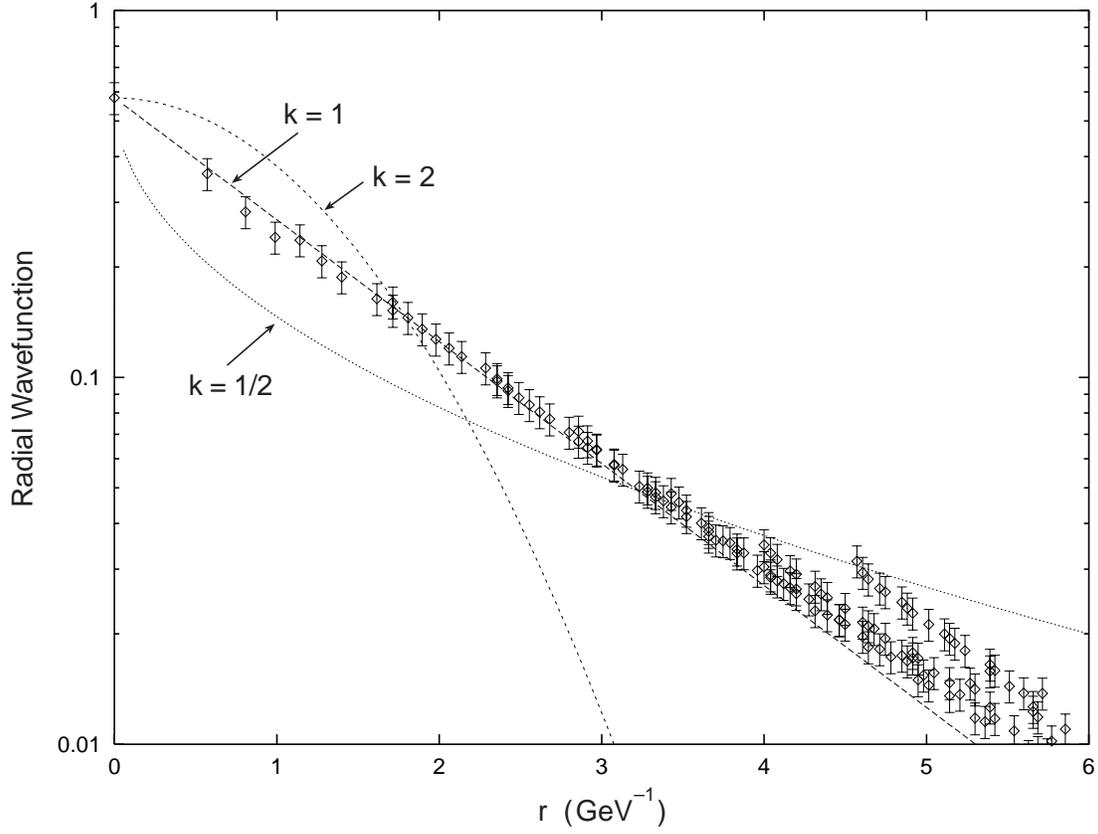,angle=270,width=6in}

\bigskip
\caption[]{The lattice data \protect{\cite{DET}}
for the heavy-light wavefunction along with
curves corresponding to the exponential parametrization (\protect{\ref{2-11}})  
with $k=0.5$, $k=1.0$,
and $k=2.0$.  It is readily seen that $k=1.0$, which corresponds to a simple  
exponential, provides a good fit.}
\label{wfcn}
\end{figure}

\begin{figure}[p]
\centering\leavevmode
\epsfig{file=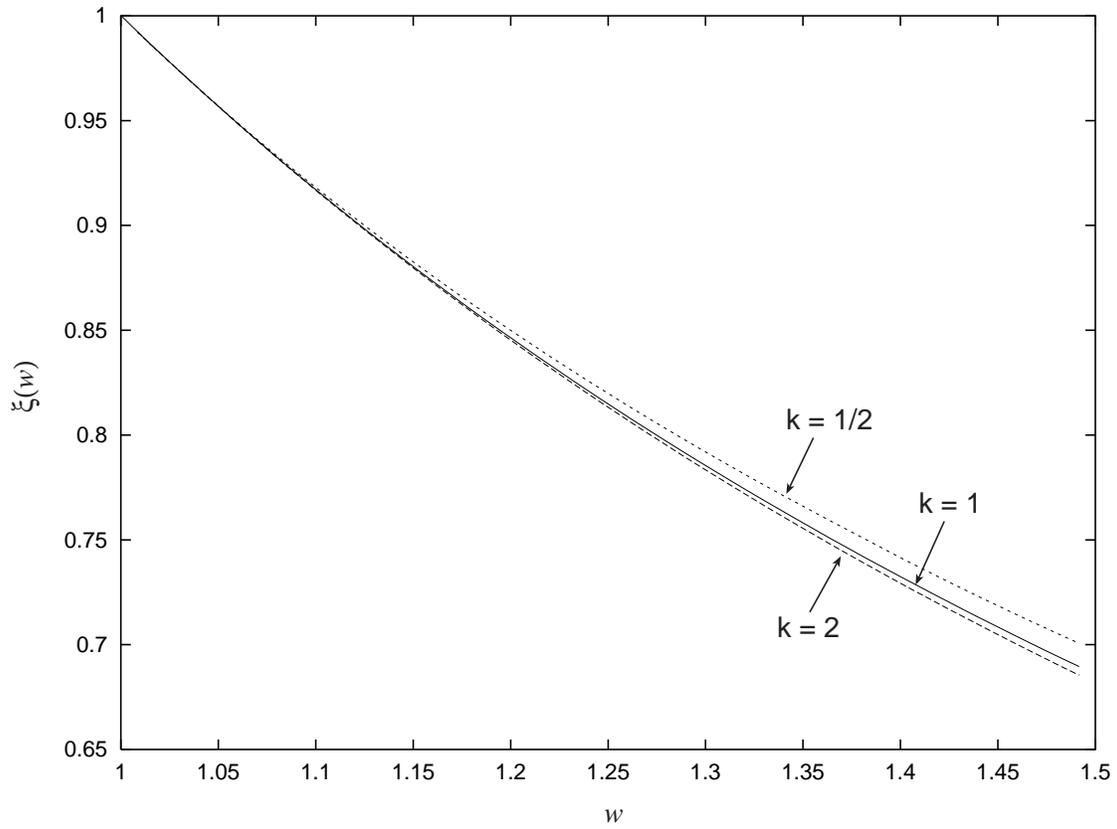,angle=270,width=6in}

\bigskip
\caption[]{Wavefunction sensitivity: The corridor for the Isgur-Wise function  
as determined using $k=0.5$
and $k=2.0$.  The solid line corresponds to $k=1.0$ All curves have
the same slope parameter, $\rho^2=0.9$.}
\label{iwk}
\end{figure}

A few years ago, the heavy-light wavefunction was evaluated in quenched
lattice simulation \cite{DET}.  The result for the ground state is shown in
Fig. \ref{wfcn}.  
One may observe that the lattice wavefunction closely resembles
a simple exponential.  We use this lattice result as a guide and
parametrize the wavefunction as \cite{OV}:
\begin{equation}\label{2-11}
f(x)=e^{-x^{k}} \,.
\end{equation}
When $k=1$ the wavefunction is a simple exponential and the lattice
simulation is closely reproduced.  In Fig. \ref{wfcn}, 
we also show that by choosing
$k=0.5$ and $k=2$ we very conservatively bracket the observed
wavefunction.  Assuming the heavy-light wavefunction is given by
(\ref{2-11}) with
\begin{equation}\label{2-12}
0.5 \leq k \leq 2 \,,
\end{equation}
the IW function is determined for all $w$ once the slope is specified.  In
Fig. \ref{iwk} we show the predicted IW function with $\rho^2=0.9$.  
The central
curve corresponds to $k=1$ with a corridor implied by the limits of
(\ref{2-12}).  We note that at $w=1.5$, which is  approximately
the largest allowed for
$B\rightarrow D^{(*)}l\nu$ decay, the corridor width is only about 0.02,
or 3\%.
This small uncertainty shows that most of
the IW shape is well-determined by the slope alone.

A particularly important special case is when $k=1$.  Straightforward
evaluation of (\ref{bigIW}) with $k=1$ in (\ref{2-11}) yields the IW
function
\begin{equation}\label{2-15}
\xi(w,\rho^2) = \frac{2(w+1)}{[w+1+(\rho^2-\frac{1}{2})(w-1)]^2} \;.
\label{our-IW}
\end{equation}
The above Isgur-Wise function will be accurate within the corridor
shown in Fig. \ref{iwk}.

\section{The Physical Form Factors}\label{FF}
An important observation from specific theoretical models \cite{review}
is that the subleading Isgur-Wise form factors which characterize
the $\frac{1}{m_Q}$
corrections are to a good approximation linear in $w$.  In addition,
the radiative QCD corrections are monotonic and vary slowly with $w$, so
that they too can be approximated linearly.
Hence each of
the $h_i(w)$ form factors (\ref{1-4}) can be written as
\begin{equation}\label{FF1}
h_i(w) \simeq [\alpha_i + \lambda_i + \mu_i(w-1)]\xi(w) \,,
\end{equation}
where both $\lambda_i$ and $\mu_i$ are small, dimensionless constants.
The advantage of this observation is that an analysis of the
experimental form factor can now be carried out without a priori
knowledge of the ${\lambda_i}$ and ${\mu_i}$ coefficients.

\begin{figure}[p]
\centering\leavevmode
\epsfig{file=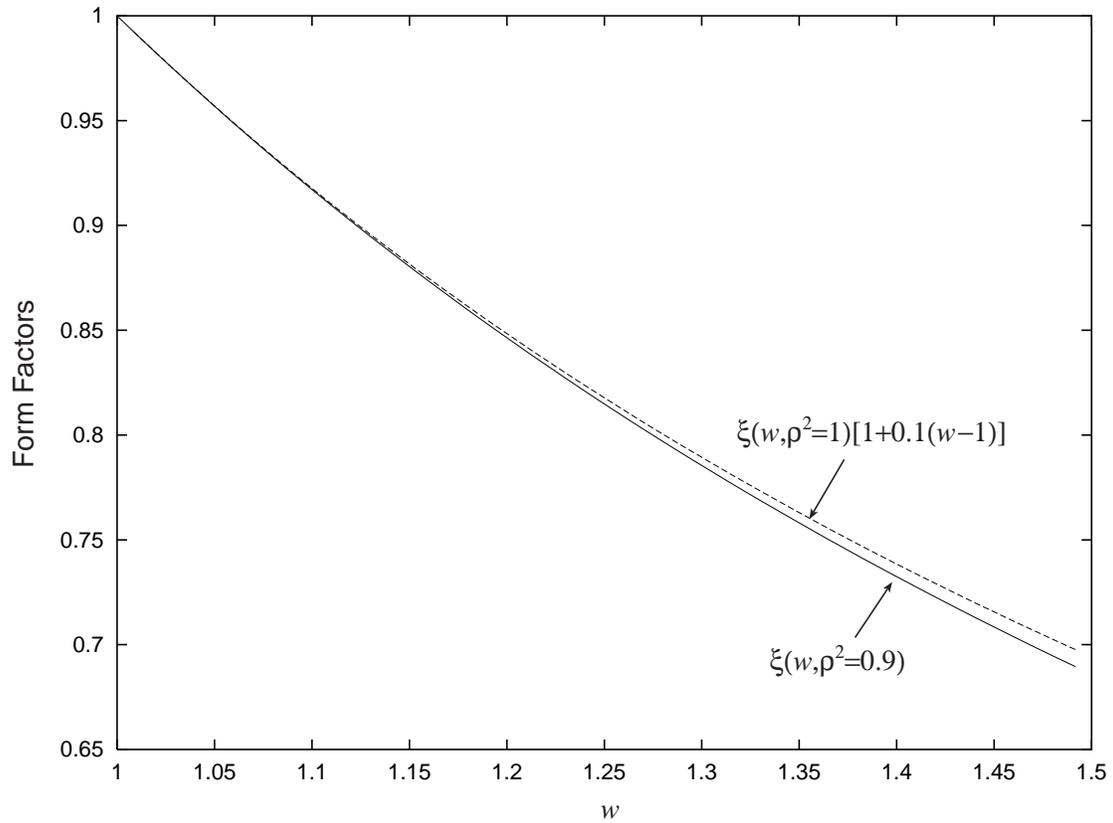,angle=270,width=6in}

\bigskip
\caption[]{Equivalence of a ``physical'' form factor to an  
Isgur-Wise function of a shifted slope. The ``shifted'' IW function $\xi(w,\rho^2=0.9)$ and the  
physical from factor approximation
$\xi(w,\rho^2=1.0)[1+0.1(w-1)]$ differ only about 2\% at $w=1.5$.
}
\label{iwcomp}
\end{figure}

To take advantage of the above, we start with the expansion (\ref{IW3})
and note that for small $(w-1)$, IW functions of slopes $\rho^2 + \Delta
\rho^2$ and $\rho^2$ are related by
\begin{equation} \label{FF3}
\xi(w,\rho^2 + \Delta\rho^2) \simeq \xi(w,\rho^2)-\Delta\rho^2(w-1) \,.
\end{equation}
Over a wider range of $w$ a more accurate expansion is
\begin{equation}\label{FF4}
\xi(w,\rho^2+\Delta\rho^2) \simeq \xi(w,\rho^2)[1-\Delta\rho^2(w-1)] \,.
\end{equation}
To illustrate the above approximation, we show in Fig. \ref{iwcomp} 
a simple physical  
form factor (with $\mu=0.1$),
\begin{equation}
h(w) = \xi(w, 1.0) [ 1 + 0.1(w-1)] \;.
\end{equation}
Also on Fig. \ref{iwcomp} 
is a shifted IW function $\xi(w, 0.9)$ with slope chosen by the  
above prescription. We note that the two form factors differ by less than  
2\% at $w=1.5$.
We now apply this result to the form factors $h_i(w)$.
Comparing (\ref{FF4}) to (\ref{FF1}) in the case
where $\alpha_i=1$ we see that
\begin{eqnarray}
h_i(w) 	&\simeq& \xi(w,\rho^2+\Delta\rho^2)+\lambda_i\xi(w,\rho^2) \;,
			\label{FF5} \\
        &\simeq& \xi(w,\rho^2+\Delta\rho^2)+\lambda_i\xi(w,\rho^2+
                   \Delta\rho^2)[1+\Delta\rho^2(w-1)] ;. \label{FF5a}
\end{eqnarray}
We drop the small product $\lambda_i \Delta\rho^2$ to obtain
\begin{equation}\label{FF6}
h_i(w) \simeq h_i(1)\xi(w,\rho_i^2) \;,
\end{equation}
with
\begin{eqnarray}\label{FF7}
\rho_i^2 &=& \rho^2 - \mu_i \;,\nonumber \\
h_i(1) &=& 1 + \lambda_i  \;.
\end{eqnarray}
We may therefore conclude that the physical form factors are nearly equivalent
to IW form factors with shifted slopes and normalizations.

As pointed out by Neubert \cite{CLN,review}, the sub-leading from
factor $\chi_1$ contributes identically to all of the $h_i$ which
remain in the heavy quark limit.  We therefore simplify the
analysis by absorbing this $\chi_1$ contribution into a new IW type
function which maintains spin-symmetry, but now has flavor dependence.
It is this slope $\rho^2$ which we refer to as the IW slope in the
following.

\begin{figure}[p]
\centering\leavevmode
\epsfig{file=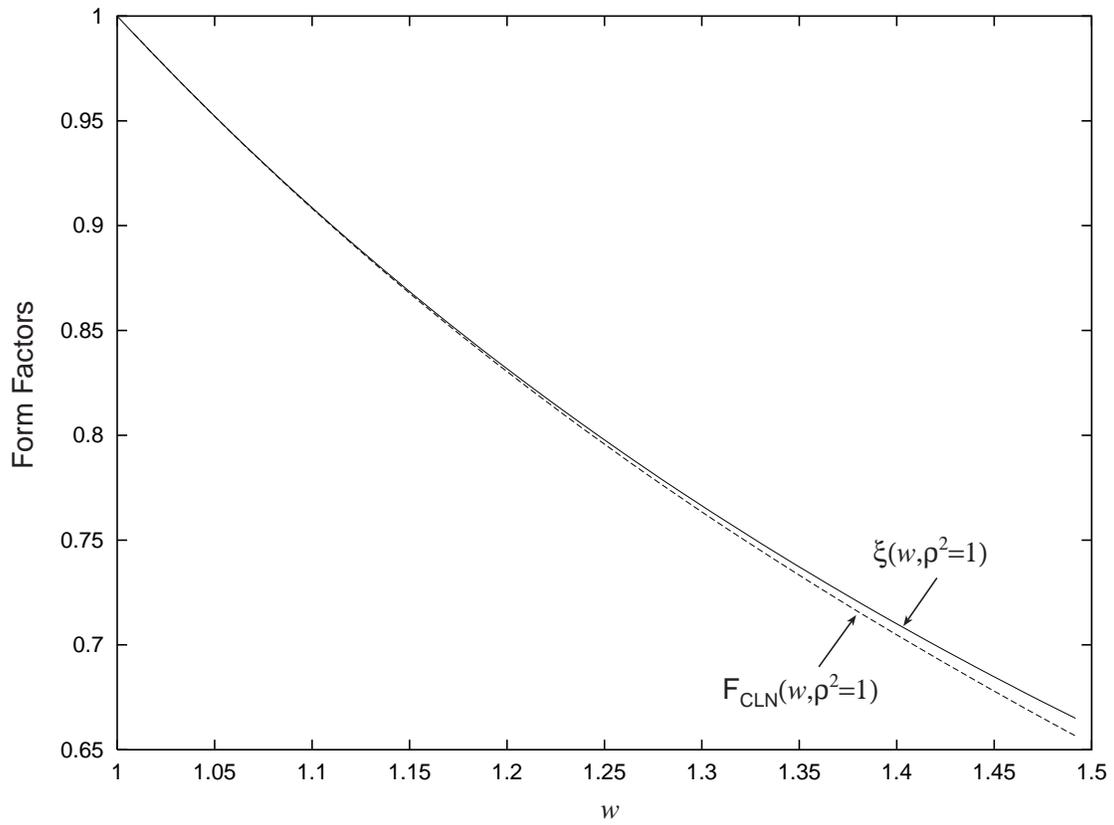,angle=270,width=6in}

\bigskip
\caption[]{The Form Factor $F(w)/F(1)$ from CLN \protect{\cite{CLN}}
along with an IW function (\protect{\ref{fcnform}})
of the same slope parameter, $\rho^2=1.0$.}
\label{ffs}
\end{figure}

The absorption of $\chi_1$ into the IW function is allowed as long
as $\chi_1$ remains small.  The evidence for this is ambiguous.
A QCD sum rule evaluation \cite{review} gives a wide range of
possible values for $\chi_1$, some of which are large.
Evidence that $\chi_1$ is small comes, indirectly, from the
dispersive analysis of Caprini, Lellouch and Neubert \cite{CLN}
where the $B \rightarrow D l \nu$ is found for a given slope.
In Fig. \ref{ffs} we show the CLN prediction for $\rho^2=1$ compared
to our $k=1$ prediction (\ref{2-15}) for the same slope.  The
two curves are nearly identical, which is expected
if $\chi_1$ and other corrections are small
and can be absorbed into an effective IW function.

\section{What Can We Learn from Experimental Form Factors?}\label{QCD}
The semileptonic $B$ decays yield definite but limited information
about QCD corrections.  It is important to understand exactly what in
a theoretical framework is being tested by experiment.  In this section,
we examine each observable quantity to see which aspect of
QCD is being tested.  We consider the semileptonic decays separately.

\subsection{\ $B\rightarrow Dl\nu$}
For $B\rightarrow Dl\bar{\nu}$, the form factor is \cite{CLN}
\begin{eqnarray}\label{FF8}
F(w) \equiv V_1(w) &=& h_+(w)+\frac{1-r}{1+r}h_-(w) \\
                   &=& \xi(w,\rho^2)[1+\lambda_D+\mu_D(w-1)+\ldots]
\end{eqnarray}
where
\begin{eqnarray}
\lambda_D &=& \lambda_{h_+} + \frac{1-r}{1+r}\lambda_{h_-} \label{lamd}
 \\
\mu_D &=& \mu_{h_+} + \frac{1-r}{1+r}\mu_{h_-} \label{mud}
\end{eqnarray}
and hence
\begin{equation}
\rho^2_D=\rho^2-\mu_D \,.
\end{equation}
The only measurable parameter here is the physical form factor slope,
$\rho_D^2$, which has been investigated at CLEO \cite{CLEOD}. If the data
improves sufficiently, the difference between $\rho_D^2$ and the  
slope of a different physical form factor (such as $F^*$)
could provide information about QCD.
To gain a rough idea of the sizes of $\lambda_D$ and $\mu_D$ we compute the  
Wilson coefficients and
substitute the QCD sum rule approximations for the actual
values of the subleading IW form factors.
The results are included in the Table II of Appendix A.

There are some directly applicable tests of HQET provided by QCD
lattice simulations.
Accurate lattice simulations are currently possible at the zero
recoil point ($w=1$).  A recent calculation \cite{FNALD} obtains:
\begin{eqnarray}
\lambda_{h_+}&=& \phantom+0.01\;,\label{h+} \\
\lambda_{h_-}&=&-0.11\;.  \label{h-}
\end{eqnarray}
In HQET, these coefficients are due to the radiative corrections and
$1/m$ corrections \cite{CLN}.  These are enumerated in Appendix A.
We observe that $\lambda_{h_+}$ has
no $1/m$ corrections as required by Luke's Theorem \cite{Luke}.  The
result from the Wilson coefficients alone of $\lambda_{h_+}=0.02$
is consistent with the lattice result (\ref{h+}).

The comparison with $\lambda_{h_-}$ of (\ref{h-}) is more interesting.
The $1/m$ correction here contains the subleading form factor $\eta$.
This form factor has been estimated by QCD sum rules \cite{LNN}, by
ISGW2 \cite{ISGW2}, and by a Salpeter equation method \cite{Vary}.
While these various calculations yield reasonably consistent results for
the subleading form factor $\chi_2(w)$, the expectations for $\eta(w)$ vary
widely.  From Appendix A we see that
\begin{equation}
\lambda_{h_-}=h_-(1)=(C_2-C_3) + \left(\frac{1}{2m_b}+\frac{1}{2m_c}\right)
(2\eta-1)
\bar{\Lambda} \,.
\end{equation}
Using the calculated Wilson coefficients and the parameters
$m_b=4.80$~GeV, $m_c=1.45$~GeV,
and $\bar{\Lambda}=0.5$~GeV we can evaluate the
previous expression to yield
\begin{equation}
\lambda_{h_-}=-0.07 + 0.12(2\eta-1) \,.
\end{equation}
The lattice result in (\ref{h-}) suggests that the second term above is
negative, i.e. $\eta<0.5$.
This prediction is not completely rigorous as the size of the effect
is of order a few percent and second-order $1/m$ corrections could be
of that order.
In addition, the lattice calculation for $h_-(1)$ has
currently been calculated only to tree level, but a 1-loop calculation
is underway \cite{FNALD}.  However, the preliminary indication
is that $\eta$ is less than the central value of 0.62 which comes
from QCD sum rules.  At the same time however, the much smaller values
of $\eta$ which come from ISGW2 and the Salpeter model are also
disfavored because they produce a result for $h_-(1)$ which is too
negative.

\subsection{\ $B\rightarrow D^* l \nu$}
In this case, there is additional information provided by the
$D^*\rightarrow D\pi$ decay distribution \cite{CLEO}.  As suggested
by CLN \cite{CLN} and followed by CLEO \cite{CLEO} this decay
can be analyzed in terms of the $h_{A_1}$ form factor and two
ratios of form factors.

The CLEO experiment, which to date has the best measurements of the
$B\rightarrow D^*$ form factor paramaters, has adopted the convention
of fitting the slope of $h_{A_1}$, $\rho^2_{A_1}$ and two form
factor ratios:
\begin{eqnarray}
R_1(w) &=& \frac{h_V(w)}{h_{A_1}(w)} \,,\nonumber \\
R_2(w) &=& \frac{h_{A_3}(w) + r^*h_{A_2}(w)}{h_{A_1}(w)} \,,
\label{Rs} \end{eqnarray}
where $r^*=\frac{m_{D^*}}{m_B} \approx 0.38$.
These ratios are expected to be nearly independent of $w$
so they are treated as constants in the fit.  The
form factor $F^*$ can be  expressed in terms of these parameters by
combining Equation (\ref{A-3}) from Appendix A 
and the definitions (\ref{Rs}).
We would also like to write $F^*(w)$ as a shifted IW function,
\begin{equation}\label{FF10}
F^*(w)=
        \xi(w,\rho^2)[1+\lambda_{D^*} +\mu_{D^*}(w-1)+\ldots] \;,
\end{equation}
and hence
\begin{equation}\label{FF11}
F^*(w)=F^*(1)\xi(w,\rho_{D^*}^2) \,,
\end{equation}
where $\rho_{D^*}^2=\rho^2-\mu_{D^*}$.
Since $|F^*|^2$ is the sum of $h_i$ contributions and kinematical factors,
the $\lambda_{D^*}$ and $\mu_{D^*}$ are the result of an expansion
about $w=1$. The result is
\begin{eqnarray}
\lambda_{D^*} &=& \lambda_{A_1} \,, \label{lamds} \\
\mu_{D^*} &=& \mu_{A_1}+\frac{1}{3}\lambda_V -\frac{1}{3(1-r^*)}
		[\lambda_{A_3}+r^*\lambda_{A_2}-r^*\lambda_{A_1}] \,.
	\label{muds}
\end{eqnarray}
As in the $B \rightarrow Dl\nu$ case, we can obtain rough
approximations of $\mu_{D^*}$, $\lambda_{D^*}$,
$\mu_{A_1}$ and $\lambda_{A_1}$.  The results are included in
Table II of  Appendix A.

The above expression (\ref{muds}) has a remarkable consequence.  The
slope difference $\rho_{D^*}^2 - \rho^2_{A_1} = \mu_{A_1}-\mu_{D^*}$
can be related to the zero-recoil quantities, $\lambda_i$ above.
It is therefore possible in the near future to test (for the first
time) the ``intercept'' quantities $\lambda_i$ by direct comparison
to experiment.  At this point only $\lambda_{A_1}$ has been
computed in lattice simulation \cite{FNALDs}, but the corresponding
values of $\lambda_{A_2}$, $\lambda_{A_3}$, and $\lambda_V$ could
be evaluated.  At present the experimental situation \cite{CLEO}
is also not precise enough:
\begin{equation}
\rho^2_{D^*}-\rho^2_{A_1}=0.08 \pm 0.15 \,,
\end{equation}
but this result can also be considerably sharpened.  We emphasize that
the above prediction (\ref{muds}) is an important check of the values
of the intercepts, $\lambda_i$, which are in turn of critical
importance in the extraction of the CKM parameter $V_{cb}$.

In the heavy quark limit,  $R_1=R_2=1$.
These ratios' deviation from unity, experimentally and theoretically,
indicates
deviation from heavy quark symmetry and, consequently,
can test theoretical predictions for the symmetry-breaking corrections.
The measurements of $R_1$ and $R_2$ (currently
with large error bars) agree with
theoretical predictions, and thus roughly confirm the predictions of
HQET.  Here we investigate the exact nature of the predictions and
their value as tests of HQET.

Treating the  form factor ratios
as constants, while reasonable theoretically and necessary for the
fit, makes the tests of symmetry-breaking corrections less
informative. $R_1(w)$ and $R_2(w)$ are essentially reduced to $R_1(1)$ and
$R_2(1)$, and because many of the corrections vanish at $w=1$, they
cannot be tested.  For example, the subleading Isgur-Wise function
$\chi_3(w)$ which has only been calculated up to the errors inherent
to QCD sum rules is completely lost in this procedure. If we apply
the expressions given in Appendix A to the various form factors
at $w=1$, eliminate those corrections which vanish at zero-recoil,
and express the results
in terms of the Wilson coefficients and
subleading IW form factors, we obtain:
\begin{eqnarray}
R_1(1) &=&
\frac{C_1(1)+\frac{1}{2m_c}\bar{\Lambda}-
          \frac{1}{2m_b}(2\eta-1)\bar{\Lambda}}
{C_1^5(1)} \\
R_2(1) &=&
\frac{C_1^5(1)+C_2^5(1)+r^*C_2^5(1)
         -\eta\bar{\Lambda}[\frac{1}{m_b}+\frac{(1+r^*)}{2m_c}]
         +\frac{4\chi_2(1)}{2m_c}(r^*-1)
	 +\bar{\Lambda}(\frac{1}{2m_b}-\frac{r^*}{2m_c})}
{C_1^5(1)}
\end{eqnarray}
If we use the theoretical estimates as rough guides to the size of
each term, we can determine what predictions are actually being tested.

\subsubsection{What Do We Learn From $R_1$?}
The subleading combination $2\eta-1$ is expected to be small on the
basis of the $\lambda_{h_-}$ considerations and is further reduced
by $\frac{1}{m_b}$.
Consequently, after evaluating the Wilson coefficients $C_i$
(see Table I of Appendix A),
$R_1$ must be greater than unity and provides a fairly direct probe of the
HQET parameter $\bar{\Lambda}$,
which is the light degrees of freedom energy to this order.

\subsubsection{What Do We Learn From $R_2$?}
In the expression for $R_2$, $\chi_2$ is generally agreed to be quite
small and $\frac{1}{2m_b}-\frac{r^*}{2m_c}\approx 0.03$.  Thus after
calculating the Wilson coefficients, a measurement of $R_2$ provides
an additional probe of the value of the subleading form factor $\eta$.
Using the Wilson coefficients in Appendix A, as well
as $m_b=4.8$~GeV, $m_c=1.45$~GeV, $r^*=0.38$ and
$\bar{\Lambda}=0.5$~GeV, the
expression for $R_2$ becomes
\begin{equation}
R_2 = 1.00 -0.34\eta(1) -0.84\chi_2(1) \,.
\end{equation}
The term containing $\eta$ is the dominant correction as $\chi_2$ is
small, so a precise measurement of $R_2$ effectively measures $\eta$.

\section{Conclusions}\label{Conc}
We have addressed the question of how to analyze semileptonic
$B$ decay while minimizing both the amount of theoretical
assumption and the number of parameters.  We first establish
that specifying the slope parameter $\rho^2$ accurately determines
the entire Isgur-Wise function.  To do so, we need only an
approximation to the light degrees of freedom wavefunction
provided by a QCD simulation.  Our result (\ref{fcnform}) is
\begin{equation}
\xi(w, \rho^2)=\frac{2(w+1)}{[w+1+(\rho^2-\frac{1}{2})(w-1)]^2} \;.
\end{equation}

The form factors $h_i(w)$ either vanish or
approach the IW function in the heavy quark limit.  These
amplitudes can be accurately written as
\begin{equation}
h_i(w)=[\alpha_i + \lambda_i +\mu_i(w-1) + \ldots]\xi(w,\rho^2) \,,
\end{equation}
where $\alpha_i$ is 0 or 1 and $\lambda_i$ and $\mu_i$ are small
dimensionless constants.  We observe that by altering the
IW slope $\rho^2$ to $\rho_i^2=\rho^2-\mu_i$ the physical form
factors can be expressed as
\begin{equation}
h_i(w)=(\alpha_i+\lambda_i)\xi(w,\rho_i^2) \,,
\end{equation}
where $\rho_i^2=\rho^2-\mu_i$.
That this works well is shown by Fig. \ref{iwcomp}.

From the above we conclude that semileptonic data are
parametrized by an intercept value $V_{cb}(1+\lambda_i)$ and an effective Isgur-Wise 
slope parameter $\rho_i^2$.  To find the CKM element $V_{cb}$
one must use a theoretical estimate of $\lambda_i$.

For $B\rightarrow D^*l\nu$ decay the parameters are
$V_{cb}(1+\lambda_{A_1})$, $\rho_{A_1}^2$, $R_1$, and $R_2$.
In this case one has a consistency condition (\ref{muds})
relating the $\lambda_i$ and the difference between
actual $D^*$ slope and $\rho^2_{A_1}$.  We
further point out that $R_1$ is nearly model-independent while
$R_2$ depends sensitively on various estimates of the subleading
from factor $\eta(1)$.  Also, the value of the form factor $h_-(w)$
at zero-recoil, $\lambda_{h_-}$, appears to offer an additional
probe of the value $\eta(1)$.

Finally, one might ask, what is the essential advantage of the slope shift  
scheme described here? The answer is, economy of parameters and a decoupling  
from theoretical assumptions. As seen in Fig. \ref{iwcomp}, 
one of the effective IW  
function slopes is equivalent to an (unknown) IW slope with subleading  
corrections. If these corrections were securely known our scheme offers no  
advantage. On the other hand, if one tried to fit both the IW slope and HQET  
$\mu$ parameters, a hopeless parameter correlation would arise.

We believe that it is best to rely as little as possible on theory and to use  
a direct phenomenological approach. As pointed out previously\cite{burd}, one  
must have some theoretical constraint on the curvature term in (\ref{1-1}). We  
show here that the shape of the form factor is specified once the slope  
parameter is given. Later, after the decay distributions have been  
parametrized, the fitted parameters can be compared to predictions.

\section*{Acknowledgements}
This research was supported in part by the U.S.~Department of Energy under Grant No.~DE-FG02-95ER40896 and in part by the University of Wisconsin Research Committee with funds granted by the Wisconsin Alumni Research Foundation.

\appendix
\section{ }
The fundamental form factors $h_i$ are defined \cite{CLN,FN} by
\begin{eqnarray}
\frac{\langle D(v')|\bar{c}\gamma^{\mu}b|B(v)\rangle}{\sqrt{m_B m_D}}
  & = & (v+v')^{\mu} h_+(w) + (v-v')^{\mu} h_-(w),  \nonumber\\
\frac{\langle D^*(v',\epsilon)|\bar{c}\gamma^{\mu}b|B(v)\rangle
}{\sqrt{m_B m_{D^*}}}
  & = &
i\epsilon^{\mu\nu\alpha\beta}\epsilon^*_{\nu}v_{\alpha}'v_{\beta}
    h_V(w), \nonumber\\
\frac{\langle D^*(v',\epsilon)|\bar{c}\gamma^{\mu}\gamma^5b|B(v)\rangle
}{\sqrt{m_B
     m_{D^*}}}
  & = &(w+1)\epsilon^{*\mu}h_{A_1}(w) - \epsilon^* \cdot v
    (v^{\mu}h_{A_2} + v'^{\mu}h_{A_3}) \label{A-1}
\end{eqnarray}

The matrix elements for $B\rightarrow Dl\bar{\nu}$ in (\ref{1-3}) can be
expressed in terms of \{$h_i$\} as
\begin{equation}\label{A-2}
F(w)=h_+(w) - \biggl(\frac{1-r}{1+r}\biggr)h_-(w)
\end{equation}
and the squared matrix element for $B\rightarrow D^*l\bar{\nu}$ in
(\ref{1-3}) as
\begin{eqnarray}
f^*|F^*(w)|^2 & = &2(w+1)(1-2wr^*+r^{*2})[(w+1)h_{A_1}^2+(w-1)h_V^2]
	\nonumber \\
   &   &+(w+1)^2[(w-r^*)h_{A_1}-(w-1)(r^*h_{A_2}+h_{A_3})]^2\,.\label{A-3}
\end{eqnarray}
The $h_i$ can, as indicated in (\ref{1-4}), be expressed by the
short-range
corrections $\beta_i$ and the $\frac{1}{m_Q}$ corrections $\gamma_i$
together as
\begin{eqnarray}
h_V(w) & = & \xi(w) [C_1+\varepsilon_c(L_2-L_5) +
      \varepsilon_b(L_1-L_4)] \nonumber \\
h_{A_3}(w) & = &\xi(w)[C_{1}^{5} + C_{3}^{5} +
                \varepsilon_c(L_2-L_3-L_5+L_6)
               +\varepsilon_b(L_1-L_4)]\nonumber \\
h_{A_2}(w) & = &\xi(w)[C_{2}^{5} + \varepsilon_c(L_3+L_6)] \nonumber \\
h_{A_1}(w) & = &\xi(w)[C_{1}^{5} + \varepsilon_c(L_2-\frac{w-1}{w+1}L_5)
	+ \varepsilon_b(L_1-\frac{w-1}{w+1}L_4)] \nonumber \\
h_+(w) & = & \xi(w)[C_1 + \frac{w+1}{2}(C_2+C_3)
                +(\varepsilon_c+\varepsilon_b)L_1 ] \nonumber \\
h_-(w) & = & \xi(w)[\frac{w+1}{2}(C_2-C_3) +
      (\varepsilon_c+\varepsilon_b)L_4] \label{A-4}
\end{eqnarray}
where $\varepsilon_{c,b} = (2m_{c,b})^{-1}$, the $C_i$ are the Wilson
coefficients --- which are discussed extensively in \cite{review} --- and the
$L_i$ are combinations of the sub-leading Isgur Wise form factors and
have been approximated by QCD sum rules
\cite{CLN}:
\begin{eqnarray}
L_1 &=& -4(w-1)\frac{\chi_2}{\xi}+12\frac{\chi_3}{\xi} \approx
	0.72(w-1)\bar{\Lambda} \nonumber \\
L_2 &=& -4\frac{\chi_3}{\xi} \approx -0.16(w-1)\bar{\Lambda}
	\nonumber \\
L_3 &=& 4\frac{\chi_2}{\xi} \approx 0.24\bar{\Lambda}
	\nonumber \\
L_4 &=& (2\eta -1)\bar{\Lambda} \approx 0.24\bar{\Lambda}
	\nonumber \\
L_5 &=& -\bar{\Lambda}  \nonumber \\
L_6 &=& -\frac{2}{w+1}(\eta+1)\bar{\Lambda} \approx
	-\frac{3.24}{w+1}\bar{\Lambda} \label{Li}
\end{eqnarray}
where $\bar{\Lambda}\approx 0.5~\rm{GeV}$ is the ``binding energy'' of a
heavy meson.

Using the detailed expressions given in \cite{CLN} and \cite{review}
and the expressions above, we can extract $\lambda_i$ and $\mu_i$ for
each form factor $h_i$ which appears in (\ref{FF1}).  They appear in
the following table.  We use the values $m_b= 4.8~\rm{GeV}$ and
$m_c=1.45~\rm{GeV}$
and calculate the Wilson coefficients at the scale
$\alpha_s(\mu)=0.12\pi$.


\begin{table}[h] \label{table3}
\caption[]{Wilson coefficient linear fits for $\alpha(\mu)=0.12\pi$ \cite{review}.
       }
\begin{tabular}{ccc}
$C_i$ & $C_i(1)$ &
                $C_i'(1)$  \\
\hline
$C_1$    & 1.13 & $-0.08$ \\
$C_1^5$  & 1.00 & $-0.04$ \\
$C_2$    & $-0.086$& 0.037 \\
$C_2^5$  & $-0.11$ & 0.038 \\
$C_3$    & $-0.019$& 0.005 \\
$C_3^5$  & 0.040 & $-0.013$\\
\end{tabular}
\end{table}

\begin{table}[h] \label{table}
\caption[]{Estimates of the $\lambda$ and $\mu$ coefficients for the various  
relevant form
        factors using the Wilson coefficients of Table~\ref{table3} and QCD  
sum rules\cite{CLN,LNN} as summarized in the Appendix.}
\begin{tabular}{ccc}
Form Factor & $\lambda_i$ &
                $\mu_i$  \\
\hline
$h_{A_1}$ & 0.004  & 0.050 \\
$F_D=V_1$ & 0.041 & 0.072  \\
$F^*={\cal F}$ & 0.004 & 0.253 \\
\end{tabular}
\end{table}

\newpage

\newpage

\newpage

\newpage

\end{document}